\documentclass[12pt]{iopart}

\newcommand{\be}{\begin{equation}}
\newcommand{\ee}{\end{equation}}
\newcommand{\bea}{\begin{eqnarray}}
\newcommand{\eea}{\end{eqnarray}}

\usepackage[dvips]{graphicx}
\usepackage{iopams}
\usepackage{amsfonts}
\usepackage{cite}
\begin{document}
\title{Quantum Quenches in Integrable Field Theories}
\author{Davide Fioretto$^1$ and Giuseppe Mussardo$^1$ $^2$}
\address{$^1$ SISSA and INFN, Sezione di Trieste, via Beirut 2/4, I-34151, Trieste, Italy}
\address{$^2$ International Centre for Theoretical Physics (ICTP), I-34151, Trieste, Italy}

\begin{abstract}
We study the non equilibrium time evolution of an integrable field theory in $1+1$ dimensions after a sudden variation of a global parameter of the Hamiltonian. For a class of quenches defined in the text, we compute the long times limit of the one point function of a local operator as a series of form factors. Even if some subtleties force us to handle this result with care, there is a  strong evidence that for long times the expectation value of any local operator can be described by a generalized Gibbs ensemble with a different effective temperature for each eigenmode.
\end{abstract}
\maketitle

\section{Introduction}
In recent years, the unitary time evolution of an extended quantum system has attracted a lot of attention. The interest of the community on this topic was spurred by novel experiments with cold atoms that have shown the coherent evolution of a quantum system in a laboratory~\cite{Greiner02,Kinoshita06, Sadler06, Hofferberth07 }.
One of the simplest and most studied setting is the so called (sudden) quantum quench. In this situation, one initially considers an extended quantum system prepared in a pure state: this can be regarded as the zero temperature ground state of an Hamiltonian ${\cal H}_0$. At t=0 the Hamiltonian is suddenly changed from ${\cal H}_0$ to ${\cal H}$ and then the system is let to evolve unitarily according to the new Hamiltonian ${\cal H}$, without any coupling to the environment. Sure enough, there is a transient regime for such an abrupt change of the Hamiltonian but the main interest is in what happens in the long time limit. A relevant issue is whether  the expectation values of local operators at $t = \infty$ can be derived by standard thermodynamical ensembles (e. g. Gibbs ensemble): if this is the case, it is said that the system thermalizes. The usual understanding is that thermalization should generally occur in extended systems but this notion has been recently challenged by experiments done on one-dimensional quasi-integrable systems, such as trapped $^{87}$Rb atoms, which do not noticeably equilibrate ever after thousands of collisions \cite{Kinoshita06}. Although the system is in a magnetic trap, this may be a manifestation of the quantum integrability of the Lieb-Liniger Hamiltonian for the atoms placed in the flat region of the trap. 
It was firstly conjectured by Rigol {\it et al.}\cite{Rigol07} that, in dealing with integrable systems, one should employ not the usual density matrix (function of the energy only) but, instead, the density matrix of a generalized Gibbs ensemble which involves all the integrals of motion $\hat{I}_l$ of the systems  
\be
\hat{\rho}_{gen}\sim e^{-\sum_l\alpha_l \,\hat{I}_l} . \label{generalized}
\ee
This conjecture has been the subject of several studies, mostly done by using specific models ~\cite{Rigol06,Cazalilla06,Gangart08,Eckstein08,Iucci09}, but also studied for initial thermal distributions~\cite{Sotiriadis09}. An important step forward was taken in the work of Barthel and Schollw\"ock \cite{Barthel08} in which they generalized a previous result by Calabrese and Cardy~\cite{Calabrese07}, proving rigourosly that for Gaussian initial states and quadratic (fermionic or bosonic) systems the conjecture does hold: a finite subsection of an infinite system indeed relaxes to a steady state described by a generalized Gibbs ensemble where the extra integrals of motions are simply the occupation numbers of each eigenmode. Moreover, they stated (postponing the proof to a future paper)  that a similar result holds also for Bethe-ansatz solvable models, even if in this case the extra integrals of motions appearing in the density matrix do not have such a simple physical interpretation.

In this paper, we tackle the issue of thermalization in integrable systems from a different point of view:  we focus our attention on continuous systems, i.e. integrable field theories in 1+1 dimensions. A reason behind this choice is that the notion of quantum integrability can be put on a firm ground only using, as criterium, the elasticity of the scattering amplitudes of the excitations of the system \cite{Sutherland04} and, indeed, this is a distinguished property of integrable field theories {\cite{Zamolodchikov79}}. Moreover, the rich analytic structure of integrable field theories could help us to obtain results that are harder to obtain with other methods. At equilibrium, quantum field theory is a very useful tool in order to describe the scaling limit of real world condensed matter systems. If such a limit exists also out of equilibrium is still an open question: in fact, after a quantum quench a large number of high energy state is populated \cite{Biroli09,Roux09}, while a field theory usually describes only the low energy excitations of the corresponding lattice model. However, studying quantum quenches in field theory is still an interesting topic, since we expect that the qualitative picture emerging from this analysis should at least give precious hints on the behavior of real world systems. The study of quantum quenches in integrable systems has also been addressed in \cite{Gritsev07,Faribault09}.

The structure of this article is as follows. In section \ref{sec_integrable}  we briefly remind the main properties of integrable field theories and we properly define the kind of  quenches we are interested in. The main result of this paper is stated in section \ref{sec_proof}, where we analyze the long time limit of the one point function of a local operator. Using a form factor expansion, we show that its asymptotic value could be obtained from a physically transparent generalized Gibbs ensemble, where the conserved charges are simply the occupation number of each eigenmode. Since the argument in section \ref{sec_proof} may appear a little abstract, in section \ref{sec_epsilon} we compute the exact time dependent one point function of the 
energy density operator of the Ising model, showing that its asymptotic value agrees with the general formula. Finally, we present our conclusions in section \ref{sec_conclusions}. In  \ref{A_proof} we gather all the technical details of our derivation.

\section{Integrable field theories and quantum quenches} \label{sec_integrable}
\subsection{Basic facts about integrable quantum field theories}\label{sub_int}
In this paper we focus our attention on integrable quantum field theories in 1+1 dimensions (for a review see e.g.~{\cite{Zamolodchikov79}, \cite{Mussardo09},\cite{Smirnov92}}). These theories are characterized  by an infinite set of local conserved currents that greatly constrain the dynamics. Indeed, only {\it elastic} scattering events can occur in these theories: no particle production is allowed and the sets of the final momenta coincide with the initial one. Moreover, due to Yang-Baxter equation, the amplitude of the $n \to n$ particle scattering event can be written in terms of the $2\to2$ scattering amplitudes. Using these remarkable properties of the scattering processes and some additional assumptions on the analytic structure of the S matrix, one is able to determine all the scattering amplitudes and, from their pole structure, the mass spectrum of the theory. For sake of clarity in the following we focus our attention on the simplest integrable field theories, i.e. those with only one excitation of mass $m$ (e.g. Ising model or Sinh-Gordon Model, where the latter is directly relevant for one-dimensional bose gases \cite{KormosMussardo1,KormosMussardo2}), even if our results could be easily extended to any integrable field theory with richer spectrum of excitations. 

A convenient basis can be constructed by the action of a creation operator $Z^\dagger (\theta)$ on a vacuum state $|0 \rangle$, where $\theta$ is the rapidity of the particle (with energy and momentum given by $E= m \cosh (\theta)$, $p=m \sinh (\theta)$ respectively). Since the theory is not free, the operators $Z(\theta)$ and $Z^\dagger(\theta)$ do not simply commute or anticommute: they satisfies instead the Zamolodchikov-Fadeev algebra 
\bea
Z(\theta_1) \, Z(\theta_2)= S(\theta_1-\theta_2) \,Z(\theta_2) \,Z(\theta_1) \,, \nonumber \\
Z^\dagger (\theta_1) \, Z^\dagger(\theta_2)= S(\theta_1-\theta_2)\, Z^\dagger(\theta_2)\, Z^\dagger(\theta_1) \, , \label{Fadeev}\\
Z(\theta_1) \, Z^\dagger(\theta_2)= S(\theta_2-\theta_1) \,Z^\dagger(\theta_2) \,Z(\theta_1) +2\,\pi\,  \delta(\theta_1-\theta_2)\, , \nonumber
\eea
where $S(\theta)$ is the two particles S matrix. Consistency of this algebra is ensured by the relationship $\overline{S} (\theta)=S(-\theta)$ and by the unitarity condition $S(\theta) S(-\theta)=\mathbf{1}$. Except for the free bosonic theory, where $S(0)=\mathbf{1}$, the $S$-matrix of all other interacting theories satisfies $S(0) =-\mathbf{1}$.   
Multi-particle states are given by acting with an ordered string of the creation operators $Z^\dagger(\theta_i)$ on a vacuum state $|0\rangle$, i.e.
\bea
|\theta_1,\dots ,\theta_n \rangle=Z^\dagger(\theta_1)\ldots Z^\dagger(\theta_n)\,|0\rangle\,, \nonumber \\ \label{basis}
\langle \theta_n , \ldots , \theta_1|=\langle 0 | \, Z(\theta_n )\ldots Z(\theta_1)= \left(|\theta_1,\dots \theta_n ,\rangle \right)^\dagger \, ,
\eea
where  
\[
\theta_1 > \theta_2 > \dots > \theta_n
\]
These orderings select a set of linearly independent vectors that form a basis in the Hilbert space.  A convenient resolution of the identity is given by
\be
\mathbf{1}=\sum_{n=0}^{+\infty}\frac 1 {n!} \int \frac {d\theta_1} {2\,\pi}\ldots  \frac{d\theta_n} {2\,\pi}\, |\theta_1,\dots ,\theta_n \rangle \langle \theta_n, \ldots, \theta_1| \, ,
\ee
where the integration is extended to all values of the rapidities.

The structure of an integrable theory is so powerful that permits to compute also the matrix elements of any local operator ${\cal O}(x)$ on the basis (\ref{basis}). Indeed, by solving a set of monodromy and recursive equations (based on the $S$-matrix), one can  determine the form factors \cite{Smirnov92}
\be
F_{\cal O} (\theta_1,\ldots ,\theta_n)=\langle 0| {\cal O}(0)|\theta_1,\ldots,\theta_n\rangle. \label{form_factors}
\ee 
Consider, for instance, the Ising field theory at $T> T_c$. This model can be described by a free fermionic theory with $S$-matrix $S=\mathbf{-1}$ and only one kind of particle of mass $ m\propto (T-T_c)$. There are three physically interesting operators in such a theory: the {\em energy operator} $\epsilon$ (that is proportional to the trace of the energy momentum tensor), the {\em spin operator} $\sigma$ (the order parameter) and the {\em disorder operator} $\mu$ related to $\sigma$ by the Kramers-Wannier duality. With a proper choice of the overall normalization, their form factors are given by the formulas~\cite{Yurov91}
\bea
\label{epsilon_def}
F_{\epsilon} (\theta_1,\ldots ,\theta_n)=\left\{
\begin{array}{c l}
-\,i\,2\,\pi\, m\, \sinh(\frac{\theta_1-\theta_2}2) & \mathrm{if  }  \,n = 2\\
0 & \mathrm{otherwise}
\end{array} \right. , \\
F_{\sigma} (\theta_1,\ldots ,\theta_n)=\left\{
\begin{array}{c l}\overline{\sigma}\,
i^{(n-1)/2} \prod_{l<m}^n \tanh(\frac {\theta_l-\theta_m} 2)  & \mathrm{if  }  \,n  \textrm{ is odd} \\
0 & \mathrm{otherwise}
\end{array} \right. , \\
F_{\mu} (\theta_1,\ldots ,\theta_n)=\left\{
\begin{array}{c l}\overline{\sigma}\,
i^{n/2} \prod_{l<m}^n \tanh(\frac {\theta_l-\theta_m} 2)  & \mathrm{if  }  \,n  \textrm{ is even} \\ \label{mu}
0 & \mathrm{otherwise}
\end{array} \right. ,
\eea
where $\overline{\sigma}=2^{\frac 1 3}\, e^{-\frac 3 4}\, A^3\, m^{\frac 1 4}$ and $A=1.282427\ldots$ is the Glasher constant\footnote[1]{Here we are using the conformal normalization of the operators, such as in the limit $x\to 0$  $\langle \phi(x,0)\phi(0,0)\rangle\sim \frac 1 {|x|^{4 \Delta}}$ where $\Delta$ is the conformal dimension of the field.}.
 
It is clear that, from formula (\ref{form_factors}), we can easily restore the $x$ and $t$ dependance of the matrix element  $ \langle 0| {\cal O}(x,t)|\theta_1,\ldots,\theta_n\rangle$, since the energy and the momentum operators are diagonal in the basis (\ref{basis}). However, it is not so obvious how to obtain from (\ref{form_factors}) the generic matrix element $ \langle \theta_n,\dots, \theta_1|{\cal O}(0) |\theta_1',\ldots ,\theta_n'\rangle$: this task can be accomplished by exploiting the crossing symmetry. If $A$ and $B$ are two sets of rapidity, we have 
\be
\langle A| {\cal O}(0)|B\rangle=\sum_{A=A_1 \cup A_2 \, ; \, B=B_1 \cup B_2} S_{A A_1} S_{B B_1} \langle A_1^+| {\cal O}(0) | B_1 \rangle \langle A_2|B_2\rangle, \label{crossing}
\ee
where the sum is over all the possible ways of splitting the sets $A$/$B$ in two subsets $A_1$/ $B_1$ and $A_2$/$B_2$ while $S_{A A_1}$ and $S_{B B_1}$ are the products of $S(\theta)$ we need to rearrange the rapidities in the proper order, namely
\bea
\langle A|=S_{A A_1} \langle A_2 A_1|\,,\\
|B\rangle=S_{B B_1} |B_1 B_2\rangle \,.
\eea
The symbol $A_1^+$ in (\ref{crossing}) denotes that each rapidity $\theta_1\ldots \theta_r$ in $A_1$ is shifted by an infinitesimal amount $\epsilon_i $ so that $\langle A_1^+| {\cal O} | B_1 \rangle  $ is simply related to the form factors (\ref{form_factors})
\be
\langle A_1^+| {\cal O}(0) | B_1 \rangle=\langle 0| {\cal O}(0) | B_1 A_1^+-i\, \pi\rangle \, . \label{cross_2}
\ee
Here comes however the tricky point.  If the $\epsilon_i$ are finite, the form factors are (for real rapidities) regular functions, while if we take the $\epsilon_i\to0$ limit (as we need to do at the end of the calculations) the form factors usually diverge, because we are in the kinematical situation in which some of the rapidities of the {\em bra} and the {\em ket} states coincide: the simplest case of this circumstance is provided by the 2-particle matrix element $\langle\theta|\mu|\theta'\rangle$ where $\mu$ is the disorder operator of the Ising model (\ref{mu}), which indeed diverges when $\theta=\theta'$.  This discussion shows that a prescription is needed for handling these kinematical divergencies. The one  proposed in \cite{LeClair_Mussardo,Balog94} consists of taking only the {\it regular} part of (\ref{cross_2}) and discarding {\em all} the terms proportional to an inverse power of $\epsilon_i$
\bea
&&\langle \theta_n, \ldots ,\theta_1| {\cal O} (0) | \theta_1',\ldots ,\theta_m'\rangle_{conn}=\label{connected}\\
&&=\textrm{Finite Parts}\left [ \lim_{\epsilon_i\to 0} \langle 0|{\cal O} (0) | \theta_1',\ldots ,\theta_m', \theta_n-i\,\pi+\epsilon_n,\ldots ,\theta_1-i\, \pi+\epsilon_1\right] \, .\nonumber
\eea
It should be stressed, however, that this prescription alone is not enough to properly take care of all the divergencies and, usually, it must be supplemented with extra corrective factors coming from the Bethe-ansatz technique \cite{LeClair_Mussardo}. Without entering in too many details, to show this aspect, consider for instance as a density matrix of the system\footnote{$\left[\hat{n} (\theta)\, , \hat{n} (\theta')\right]=0$, so we do not have ordering problems in the exponential.} 
\be
\hat{\rho}_{\lambda}=\frac{\exp\left(-\int d \theta \lambda (\theta) \,\hat{n} (\theta)\right)}{Z} , \label{ensemble}
\ee
where $\hat{n}(\theta)=Z^\dagger(\theta) Z(\theta)$  and $\lambda(\theta)$ is an appropriate function of $\theta$. Notice 
that, if $\lambda (\theta)=\frac 1 T\, m\, \cosh(\theta)$ the density matrix (\ref{ensemble}) describes the familiar canonical ensemble,  while, in the more general case, it can be associated to the  generalized Gibbs ensemble (\ref{generalized}): indeed, in an integrable field theory the (infinite) set of local continuity equations determines the conserved charges $\hat{Q}_s$ labeled by an integer index $s$  \cite{Zamolodchikov79}, and each charge $\hat{Q}_s$ can be written as
 \be
 \hat{Q}_s= q_s\, \int d \theta \,e^{s \theta}\,\hat{n} (\theta),  
\ee  
where $q_s$ is a constant. Obviously, it would be nice to write down the average of a local operator $\cal O$ w.r.t. the density matrix ($\ref{ensemble}$) in term of these form factors. But, if we do it applying the prescription (\ref{connected}) alone we end up with the result
\bea
&&\!\!\!\!\!\langle O \rangle_{\hat{\rho}}=Tr\left( \hat{\rho}\, O(0) \right)= \;\;\;\;\;\;\;\;\;\;  \;\;\;\;\;\;\;\;\;\; \;\;\;\;\;\;\;\;\;\; \;\;\;\;\;\;\;\;\;\; \;\;\;\;\;\;\;   \mathbf{! \, WRONG \,!} \label{wrong}\\
&&\!\!\!\!\!\!=\sum_{n=0}^{+\infty} \frac 1 {n!} \int \frac{d\theta_1 \dots d\theta_n} {(2 \pi)^n} \prod_{i=1}^n \left[\frac {e^{-\lambda (\theta_i)}}{1-S(0)\, e^{-\lambda (\theta_i)}} \right]\langle \theta_n ,\dots ,\theta_1|O(0)|\theta_1,\ldots,\theta_n\rangle_{conn}\nonumber\, ,
\eea 
which is wrong since it does not agree with the thermodynamic Bethe ansatz. It was firstly conjectured by LeClair and Mussardo~\cite{LeClair_Mussardo} that the correct expression is instead
\bea
&&\!\!\!\!\!\langle O \rangle_{\hat{\rho}}=Tr\left( \hat{\rho} \, O(0) \right)= \label{average}\\
&&\!\!\!\!\!\!=\sum_{n=0}^{+\infty} \frac 1 {n!} \int \frac{d\theta_1 \dots d\theta_n} {(2 \pi)^n} \prod_{i=1}^n \left[\frac {e^{-\tilde{\lambda} (\theta_i)}}{1-S(0)\, e^{-{\tilde\lambda} (\theta_i)}} \right]\langle \theta_n, \dots ,\theta_1|O(0)|\theta_1,\ldots,\theta_n\rangle_{conn}\nonumber\, ,
\eea
where the $\tilde{\lambda}$ are dressed according to the integral equation
\be
\tilde{\lambda}(\theta)=\lambda(\theta)-\int \frac{d \theta'}{2\,\pi} \, \varphi(\theta-\theta') \log [1+e^{-\tilde{\lambda}(\theta')} ], \label{dressing}
\ee 
$\varphi$ being the derivative of the phase shift
\be
\varphi(\theta)=-i \frac d {d \theta} \log( S(\theta)) .
\ee 
Originally,  (\ref{average}) was only an educated guess but its correctness was confirmed by subsequent checks~\cite{Saleur00,Lukyanov01,Mussardo01}. Even if presently there is still no complete proof of the exactness of (\ref{average}), a step forward its rigorous derivation has been taken in the recent papers by Pozsgay and Takacs~\cite{Pozsgay_Takacs08_1,Pozsgay_Takacs08_2,Pozsgay09}, who showed  that (\ref{average}) might  be obtained in a rigorous way by putting the system in a finite volume, thus discretizing the rapidities to regularize the kinematical singularities, and taking the infinite volume limit only at the end of the calculations. 

\subsection{Integrable boundary condition and quantum quenches} 
Our aim is to study the time evolution of the expectation value of a local operator ${\cal{O}} (x)$ on a pure state $|B\rangle$ that is not an eigenstate of the integrable Hamiltonian $\cal H$, i. e.
\be
\langle O (x,t) \rangle_{B} =\frac{\langle B| e^{i\, {\cal H}\, t} {\cal O}(x) e^{-i\, {\cal H}\, t}|B\rangle}{\langle B|B\rangle}.
\ee
As shown by Calabrese and Cardy \cite{Calabrese06,Calabrese07}, performing a Wick rotation, this dynamical problem can be mapped into a statistical  problem defined in a strip geometry, where  the initial state $|B\rangle$ plays the role of a boundary condition.  In an integrable field theory, the most natural boundary states are the ones that do not spoil the integral of motions. These integrable boundary states were originally studied by Ghoshal and Zamolodchikov in \cite{Ghoshal94} and their properties can be summarized as follows: these states have a peculiar form, given by a coherent superposition of pairs of equal and opposite rapidity (Cooper pairs)\footnote{Here, for simplicity, we ignore the presence of additional zero-rapidity terms in the expression of $| B \rangle$.}
\be
|B\rangle=\exp \left[\frac 1 2 \int_{-\infty}^{+\infty} \frac{d \theta}{2\, \pi} K^{a b}(\theta) Z^\dagger_a(-\theta) \, Z^\dagger_b(\theta))\right]|0\rangle, \label{integrable_boundary}
\ee
where the indices $a$ and $b$ run over all the particles. The amplitude $ K^{a b}(\theta)$ satisfies a set of equations that depends on the S matrix (such as boundary Yang Baxter, boundary unitarity and crossing equations): the different solutions of these equations provide the set of integrable boundary conditions of the theory in question. As an explicit example, let's consider once again the Ising model, in which there are three integrable boundary conditions associated to the amplitudes
\bea
\begin{array}{l l}
K_{fix} (\theta)=i \tanh(\frac{\theta} 2) & \textrm{Fixed Boundary Condition}\\
K_{free} (\theta)=-i \coth(\frac{\theta} 2) & \textrm{Free Boundary Condition}\\
K_h (\theta)=i \tanh(\frac{\theta} 2) \frac{k+\cosh(\theta)}{k-\cosh(\theta)} &  \textrm{Magnetic Boundary Condition,}
\end{array}
\label{Ising_boundary}
\eea 
where $k=1-\frac{h^2}{2 m} $ and $h$ is a real number (the boundary magnetic field). Clearly, the magnetic boundary condition interpolates between the free and the fixed ones. So, we see that only particular choices of the amplitude $K$ give us an integrable boundary state. 

Our interest in the following is to study a particular class of quenches, i.e. those where the initial state is expressed by a {\em coherent superposition of particle pairs}, as the one given in eqn.(\ref{integrable_boundary}). Among these quenches there are, of course, the integrable boundary states but the whole class of these quenches is actually larger than the genuine integrable ones: indeed, for theories with only one particle, it suffices that the amplitude $K(\theta)$ satisfies the  ``boundary cross- unitarity equation''\footnote{For theories with many particles, $K_{ab}(\theta)$ must also solve an appropriate boundary Yang-Baxter equation to ensure commutativity of the various terms in the exponential expression (\ref{integrable_boundary}).}
\be
K(\theta)=S(2\theta) K(-\theta) ,\label{cross_uni} \,\,\,.
\ee
This equation simply comes from the request of invariance of the expression (\ref{integrable_boundary}) under a change of variable $\theta\to - \theta$. 

Notice that, at least for a free fermionic theory as the Ising model, a boundary state of the form (\ref{integrable_boundary}) has a very simple meaning within the theory of quench processes. Suppose in fact that ${\cal H}_0$ is a free fermionic theory, with its ground state $|B\rangle$ identified by the condition $Z_0(\theta) |B\rangle=0$, where $Z_0(\theta)$ is the annihilation operator related to ${\cal H}_0$. If, at $t=0$, the Hamiltonian is suddenly changed from ${\cal H}_0$ to a new almost free Hamiltonian ${\cal H}$ (for instance, making a quench of the mass from $m_0$ to $m$), the new sets of creation/annihilation operators $Z^\dagger, Z$ will be related to the initial ones by a Bogoliubov transformation of the form  
\be
Z_0(\theta)\sim \left[ Z(\theta)-f(-\theta) Z^\dagger(-\theta) \right],
\ee
Hence, the algebraic equation that identifies the boundary state in terms of the new creation/annihilation operators is given by 
\be
Z_0(\theta) \,|B \rangle \,=0 \Longrightarrow \,\left[Z(\theta)-f(-\theta) Z^\dagger(-\theta)\right]\,| B \rangle \,=\,0
\ee 
whose solution is 
 \be
 |B\rangle=\exp \left[\frac 1 2 \int_{-\infty}^{+\infty} \frac{d \theta}{2\, \pi} f(\theta) Z^\dagger(-\theta) \, Z^\dagger(\theta))\right]|0\rangle .
\ee
Summarizing, in the following we consider those quantum quenches where the initial state is given by a superposition of Cooper pairs according to \ref{integrable_boundary}), where the amplitude $K$ is a (regular) function that satisfy (\ref{cross_uni}). However the boundary state as written in  (\ref{integrable_boundary}) is typically not normalizable.  Indeed, if the amplitude $K$ does not goes to zero for large rapidities (as in eq.(\ref{Ising_boundary})), it means that we are exciting modes with arbitrary high energy, hence the state has an unphysical divergent energy density. So, as done for the critical systems 
\cite{Calabrese06,Calabrese07}, it is therefore convenient to introduce an extrapolation time $\tau_0$ that make the norm of the state (\ref{integrable_boundary}) finite: this parameter plays the role of an ultraviolet cut-off, in any case present in physical systems. So, we are interested in 
\be
\langle O (x,t) \rangle_{\tilde{B}} =\frac{\langle \tilde{B}| e^{i\, {\cal H}\, t} {\cal O}(x) e^{-i\, {\cal H}\, t}|\tilde{B}\rangle}{\langle \tilde{B}|\tilde{B}\rangle}, \label{starting}
\ee
where
\be
|\tilde{B}\rangle=\e^{-{\cal H} \, \tau_0} |B\rangle=\exp \left[\frac 1 2 \int_{-\infty}^{+\infty} \frac{d \theta}{2\, \pi} G (\theta) Z^\dagger(-\theta) \, Z^\dagger(\theta))\right]|0\rangle , \label{boundary}
\ee
and $G(\theta)= e^{-2\, m\,\tau_0\, \cosh(\theta)} K(\theta)$. Notice that the extrapolation time does not spoil the basic equation of the boundary state, namely also the new amplitude $G(\theta)$ satisfies (\ref{cross_uni})
\be
G(\theta)=S(2\theta)\, G(-\theta)\,\,\, \textrm{and}\,\,\,\overline{G}(\theta)=S(-2\theta)\, \overline{G}(-\theta)\, . \label{cross_uni2}
\ee

\section{Long time behavior} \label{sec_proof}
In this section, we are concerned with the computation of 
\be
\overline{\cal{O}}=\lim_{t\to+\infty}  \langle O (x,t) \rangle_{\tilde{B}}\,\,, \label{limit}
\ee
and in the following we will show that $\overline{\cal O} $ can be expressed as an average over a density matrix like (\ref{ensemble}). At first sight, it seems very unlikely that we can do so: our boundary state (\ref{boundary}) has a very peculiar structure, since it is the superposition of pairs of opposite rapidity, while in an average over an ensemble (\ref{average}) there is no sign of such a structure. How comes that the system retains no memory of this pair structure in the long time limit?  

First of all, a trivial remark: since $|\tilde{B}\rangle $ is translational invariant, $\langle O (x,t) \rangle_{\tilde{B}}$ do not depend on $x$ and therefore, from now on, we will set $x=0$. In principle, it is quite clear what we have to do in order to compute (\ref{limit}): first we expand the exponential in (\ref{boundary}) and, taking into account that the Hamiltonian is diagonal in the particle basis, we thus arrive to the double sum
\bea
&&\!\!\!\!\!\!\!\!\langle O (x,t) \rangle_{\tilde{B}} =\frac 1 {\langle \tilde{B}|\tilde{B}\rangle} \sum_{n,l=0}^{+\infty} \frac 1 {n!\, l! }\int \frac{d\theta_1 \ldots d\theta_n}{(4\, \pi)^n} \, \frac{d\theta'_1 \ldots d\theta'_l}{(4\, \pi)^l} \exp\left[2\,i\,t(E_n(\theta)-E_l(\theta')) \right]\nonumber \\
&&\!\!\!\!\!\!\!\!\left[ \prod_{i=1}^n \overline{G}(\theta_i)\right]  \left[ \prod_{j=1}^l G(\theta'_j)\right]  \langle \theta_n,-\theta_n,\ldots \theta_1,-\theta_1|{\cal O}|-\theta'_1,\theta'_1,\ldots-\theta'_l,\theta'_l\rangle\,, \label{double}
\eea
where we used the short-hand
\be
E_n(\theta)=m \sum_{i=1}^n \cosh(\theta_i)\,.
\ee
However it is difficult to compute the long time limit directly from (\ref{double}). The reason is that the matrix element in (\ref{double}) are not regular functions (rather they have delta-like contributions) and, in such a case, we cannot apply a stationary phase argument. In order to isolate the singular parts, we have to employ eqn.(\ref{crossing}). Consider, for instance, the term with $n=l=1$ in the numerator of (\ref{double}),  i.e. 
\be
\int \frac{d\theta_1}{4\,\pi} \frac{d\theta'_1}{4\,\pi} \,\exp\left[2\,i\,t(E_1(\theta)-E_1(\theta')) \right]  \overline{G}(\theta_1)G(\theta'_1)  \langle\theta_1,-\theta_1|{\cal O}|-\theta'_1,\theta'_1\rangle \, .
\ee
Applying the crossing relation (\ref{crossing}), we can recast this term as
\bea
&&\!\!\!\!\!\!\!\!\!\! \int \frac{d\theta_1}{4\,\pi} \frac{d\theta'_1}{4\,\pi} \,\exp\left[2\,i\,t(E_1(\theta)-E_1(\theta')) \right]  \overline{G}(\theta_1)G(\theta'_1)  \langle\theta_1^+,-\theta_1^+|{\cal O}|-\theta'_1,\theta'_1\rangle +\label{num}\\
&&\!\!\!\!\!\!\!\!\!\!  + \int \frac {d\theta_1}{2\, \pi} |G(\theta_1)|^2 \langle \theta_1^+|{\cal O}|\theta_1\rangle+\langle 0| {\cal O}| 0 \rangle \int \frac{d\theta_1}{4\,\pi} \frac{d\theta'_1}{4\,\pi} \, \overline{G}(\theta_1)G(\theta'_1)  \langle\theta_1,-\theta_1|-\theta'_1,\theta'_1\rangle \, ,\nonumber
\eea 
where we make use of the symmetry properties (\ref{cross_uni2})
of $G$. Actually, the inner product $ \langle\theta_1,-\theta_1
|-\theta'_1,\theta'_1\rangle$ above is divergent in the infinite volume limit and it should be regularized by putting the system in a box of length $L$, obtaining
\be
(\delta(\theta-\theta'))^2=\frac{m \, L}{2\, \pi} \cosh(\theta)\delta(\theta-\theta') .
\ee
This, however, is not an important contribution since, as shown in \ref{A_proof}, all these inner products cancel out with the denominator of ({\ref{double}). What is really crucial is that, apart from these infinite volume divergencies that we can easily regularize, the integrands in  (\ref{num}) are all well-behaved functions:  hence we can now easily take the infinite time limit $t \to+\infty$, so that (\ref{num}) simply becomes
\be
\!\!\!\!\!\!\!\!\! \int \frac {d\theta_1}{2\, \pi} |G(\theta_1)|^2 \langle \theta_1^+|{\cal O}|\theta_1\rangle+\langle 0| {\cal O}| 0 \rangle \int \frac{d\theta_1}{4\,\pi} \frac{d\theta'_1}{4\,\pi} \, \overline{G}(\theta_1)G(\theta'_1)  \langle\theta_1,-\theta_1|-\theta'_1,\theta'_1\rangle \, 
\ee
because the first term in (\ref{num}) vanishes for the fast oscillation of its integrand. 

In the light of this example, the strategy to compute the expectation values of local operators can be stated as follows.
\begin{enumerate} 
\item We first expand the exponential in the numerator of (\ref{boundary}), ending up with the double sum (\ref{double}). 
\item Then, we use (\ref{crossing}) in order to isolate the delta-like terms and, after having done that, we take the infinite time limit, where all  terms that explicitly depend on time go to zero, due to the fast oscillation of the integrand. This is a simple consequence of the stationary phase argument, that can also be seen in the following way. If the infinite time limit exists (and the stationary phase argument assures us that it does exist), then it must coincide with the temporal average
\be
\!\!\!\overline{\cal{O}}=\lim_{t\to+\infty}  \langle {\cal O} (x,t) \rangle_{\tilde{B}}=\lim_{T\to+\infty}\frac 1 T \int_0^T dt' \langle O (x,t') \rangle_{\tilde{B}}\,\,.
\ee
We know that  the time-dependent part of the numerator of (\ref{double}) consists of a sum of terms as
\be
\!\!\!\!\!\!\!\!\!\!\!\!\!\!\!\! \int d\theta_1 \ldots d\theta_n \, d\theta'_1 \ldots d\theta'_l \exp\left[2\,i\,t(E_n(\theta)-E_l(\theta')) \right]F(\theta_1 \ldots \theta_n,\theta'_1 \ldots \theta'_l)\, ,
\ee
where $F$ is the  regular function obtained by applying (\ref{crossing}). So, the only contributions to the infinite time limit comes from the region $E_n(\theta)=E_l(\theta')$, whose Lebesgue measure is zero, so  the integral goes to zero since $F$ has no delta-like term. 

\end{enumerate}
With these steps in mind, it is a nice combinatorial exercise to show that $ \overline{\cal{O}}$ can be finally expressed as
 \be
\!\!\!\!\!\! \!\!\!\!\!\! \!\!\!\!\! \overline{\cal{O}}=\sum_{n=0}^{+\infty} \frac 1 {n!} \int 
\prod_{i=1}^n 
\frac{d\theta_i}{(2 \pi)}\,
\left[\frac {|G(\theta_i)|^2}{1-S(0)\, |G(\theta_i)|^2} \right]\langle \theta_n+i \,\epsilon_n \dots \theta_1+i \,\epsilon_1|O(0)|\theta_1\ldots\theta_n\rangle \, . \label{infinite_time0}
 \ee
In order to  keep our exposition clear, we will present the details of the combinatorics behind (\ref{infinite_time0}) in the \ref{A_proof}.  However, (\ref{infinite_time0}) is still a meaningless expression, since we have to regularize it in a proper way. One way to do it is by analogy with LeClair and Mussardo formula, discussed in section \ref{sub_int}. When we perform an average over a density matrix (\ref{ensemble}), we end up with the following expression
\be
\!\!\!\!\!\! \!\!\!\!\!\! \!\!\!\!\! \overline{\cal{O}}=\sum_{n=0}^{+\infty} \frac 1 {n!} \int 
\prod_{i=1}^n 
\frac{d\theta_i}{(2 \pi)}\,
\left[\frac {e^{-\lambda(\theta)}}{1-S(0)\, e^{-\lambda(\theta)}} \right]\langle \theta_n+i \,\epsilon_n \dots \theta_1+i \,\epsilon_1|O(0)|\theta_1\ldots\theta_n\rangle \, . 
 \ee
 LeClair and Mussardo suggested that the proper way to regularize the $\epsilon_i\to 0$ limit of this expression is to take the the connected part of the form factors (\ref{connected}) and to dress $\lambda(\theta)$ according to the integral equation (\ref{dressing}). This regularization scheme holds for every function $\lambda(\theta)$. The situation is the same in equation (\ref{infinite_time0}), with $|G(\theta)|^2$ that plays the role of $e^{-\lambda(\theta)}$.  So, the natural way of regularize ({\ref{infinite_time0}) lead us to
 \be
\!\!\!\!\!\! \!\!\!\!\!\! \!\!\!\!\! \overline{\cal{O}}=\sum_{n=0}^{+\infty} \frac 1 {n!} \int 
\prod_{i=1}^n 
\frac{d\theta_i}{(2 \pi)}\,
\left[\frac {|\tilde{G}(\theta_i)|^2}{1-S(0)\, |\tilde{G}(\theta_i)|^2} \right]\langle \theta_n\dots \theta_1|O(0)|\theta_1\ldots\theta_n\rangle_{conn} \, ,
 \label{infinite_time1}
 \ee 
 where $|\tilde{G(\theta)}|^2$ is dressed in the same way as the term 
  $e^{-\tilde{\lambda}(\theta)}$ entering the thermodynamic Bethe ansatz 
\be
|\tilde{G}(\theta)|^2=|G(\theta)|^2\,\exp\left[\int \frac{d \theta'}{2\,\pi} \, \varphi(\theta-\theta') \log [1+|\tilde{G}(\theta)|^2 ] \right] \, . \label{dressingG}
\ee 
The above dressing formula is based on the LeClair and Mussardo conjecture (it has actually the same mathematical structure) and on the possibility to  exchange the $\epsilon_i\to0$ and $t\to+\infty$ limits.  Assuming that such a regularization scheme is indeed correct, it turns out that that the long time limit of (\ref{starting}) could be described by a generalized Gibbs ensemble (\ref{ensemble}), where the constant of motions are simply given by the occupation number $\hat{n}(\theta)$ and the function $\lambda(\theta)$ is fixed by the conditions 
\be
\frac{\langle \tilde{B}|  \hat{n}(\theta)|\tilde{B}\rangle}{\langle \tilde{B}|\tilde{B}\rangle}=Tr\left(\hat{\rho}_{\lambda} \,\hat{n}(\theta)\right) ,
\ee
 thus proving Rigol {et al.}'s conjecture for integrable field theory. If we look at our starting point (\ref{starting}), this result is quite unexpected: when we expanded the exponential in (\ref{starting}) we had a double sum, and it was only thanks to the infinite time limit that we could rewrite it as a single summation. Moreover, while the boundary state (\ref{boundary}) is formed by pairs of particles with opposite rapidity, this feature is completely lost in the final expression (\ref{infinite_time1}).
\section{The simplest example} \label{sec_epsilon}
It is instructive to see how the general ideas of the previous section apply in the simplest case provided  
by the one point function of the $\epsilon$ operator of the Ising model (\ref{epsilon_def}). Indeed, for this operator we can calculate exactly its one point function for any time with elementary techniques.  From the form factors (\ref{epsilon_def}), it follows that the operator $\epsilon$ is a quadratic form in the creation - annhilation operators
\bea
&&\epsilon(0)=\int \frac{d\beta_1}{2\,\pi} \frac{d\beta_1}{2\,\pi}  \bigg \{2\,\pi \, m \cosh \left(\frac{\beta_1-\beta_2}2\right)Z^\dagger(\beta_1)  Z(\beta_2)+ \nonumber \\
&&+\left[ i \,\pi\, m \sinh \left(\frac{\beta_1-\beta_2} 2\right)  \right] \left[Z(\beta_1)  Z(\beta_2)+Z^\dagger(\beta_1)  Z^\dagger(\beta_2) \right] \bigg \} .
\eea
Since the theory is free, we can easily calculate the expectation value of binomials of the creation-annihilation operators on the boundary states (introducing, for instance, a generating functional)
and we have that
\be
\frac{\langle \tilde{B} (t)| Z^\dagger(\beta_1)\, Z(\beta_2) |\tilde{B}(t)\rangle}{\langle \tilde{B}|\tilde{B}\rangle}= 2 \, \pi \,\delta(\beta_1-\beta_2)\,\, \frac{|G(\beta_1)|^2} {1+|G(\beta_1)|^2},
\ee
\be
\frac{\langle \tilde{B} (t)| Z(\beta_1)\, Z(\beta_2) |\tilde{B}(t)\rangle}{\langle \tilde{B}|\tilde{B}\rangle}= 2 \, \pi \,\delta(\beta_1+\beta_2)\,\, \frac{G(\beta_1)e^{-2\, i\, t\, E_1(\beta_1)}} {1+|G(\beta_1)|^2},
\ee
Hence
\bea
\langle \epsilon (x,t) \rangle_{\tilde{B}}=&&+ 2\,\pi \, m \int \frac {d \theta} {2 \pi}  \frac{|G(\theta)|^2} {1+|G(\theta)|^2}+ \label{exact} \\
&&-2\,\pi \, m \int \frac {d \theta} {2 \pi}  \frac{\sinh(\theta)\, Im\left[G(\theta)\, \exp\left(-2\, i\,t E_1 (\theta) \right)  \right] } {1+|G(\theta)|^2}\, . \nonumber
\eea 
The results (\ref{exact}) has all the features we expect to hold in the general case. First of all, the time dependent part goes to zero as a consequence of the fast oscillation of the integrand. The long time asymptotic value obviously agrees with our general result: in this case, the structure of the operator is so simple that the entire sum reduces to a single term. More important, since we are able to calculate exactly the time dependence, we can also estimate the approach to the $t\to +\infty$ limit value using a stationary phase approximation. It turns out that (\ref{exact}) approaches its asymptotic value as an inverse power law, in contrast with the exponential decay of the massless case\cite{Calabrese06} \cite{Calabrese07}. Heuristically, the different behavior of the two cases (massive and massless) can be understood in the following way. We know that the relaxation to the stationary value is due to the speed of the quasiparticle excitations: in a massless theory, all the quasiparticles move at the fastest possible velocity, so we can expect that the decay to the limit value is the fastest possible (exponential decay),  while, in a massive theory, the time decay has to be slower for the spread of the velocity distribution of the quasi-particles. It may be interesting to compare this continuum formula (\ref{exact}) with the one point function of the transverse magnetization of the quantum Ising model\cite{Barouch70}:  a quench of the transverse magnetization in the lattice corresponds to a mass quench in the continuum. In both cases there are oscillations modulated by a power law with exponent $-\frac 3 2$. However, it must be noted that  the long time leading behavior of the transverse magnetization on the lattice is not only dictated by the small momenta, and therefore it is not a surprise that the continuum theory is unable to capture all the features of the lattice model.

\section{Conclusions} \label{sec_conclusions}
In this paper we studied the quantum quenches in the context of integrable field theories in 1+1 dimension.  We considered a specific class of quenches, those whose initial state is a coherent superposition of Cooper pairs  (\ref{boundary}), a structure that is in agreement with the integrability of the theory in the bulk. For such quenches, we analyzed the one point function of a local operator and argued that, in the long time limit, it can be neatly expressed in terms of the formulas (\ref{infinite_time1} ) and (\ref{dressingG}): the final result employs a series over the connected form factors (\ref{connected}), which are finite expressions. 

As discussed extensively in the text, our final result must be presently considered only as a solid conjecture since it is based on a regularization scheme of the kinematical singularities of the original form factors. This regularization is closely related to the one used in LeClair and Mussardo formula for the one point function at finite temperature \cite{LeClair_Mussardo} -- formula that has successfully passed many checks although it is not yet rigorously proved. In this respect, we have performed an additional but independent test of our result for a particular one point function of the Ising model.
If the full correctness of eqn. (\ref{infinite_time1}) will be confirmed by a future rigorous analysis, it follows that the asymptotic value of  the one point function of a local operator could be computed as an average over a density matrix (\ref{ensemble}) with a temperature different for each eigenmode: in short, our result is a strong evidence that Rigol {\it et al.}'s conjecture does hold for integrable field theories.

In this paper we limited ourselves to integrable theory with only one massive particle but our results could be extended to theories with more than one particle without any physical difficulty, although with a combinatorics much more involved.  Finally, our results seems to suggest that in a generic massive integrable theory the decay towards the asymptotic value is dictated by a power law (in contrast with the conformal exponential decay) whose leading exponent can be determined by a stationary phase approximation.

\ack We would like to thank G. Brandino, M. Burrello, P. Calabrese, E. Canovi, G. Gori, T. Macr\`i, D. Rossini, A. Silva and S. Sotiriadis for many fruitful discussions. This work is supported by the grants INSTANS (from ESF) and 2007JHLPEZ (from MIUR).

\appendix
\section{A diagrammatic proof of  eq. \ref{infinite_time0}} \label{A_proof}
In this appendix we would like to show that, for $t\to +\infty$,
\bea
&&\frac{\langle \tilde{B}|  {\cal O}(x,t)|\tilde{B}\rangle}{\langle \tilde{B}|\tilde{B}\rangle} \to \label{A_starting} \\
&&\to \sum_{n=0}^{+\infty} \frac 1 {n!} \int \frac{d\theta_1 \dots d\theta_n} {(2 \pi)^n} \prod_{i=1}^n \left[\frac {|G(\theta_i)|^2}{1-S(0)\, |G(\theta_i)|^2} \right]\langle \theta_n^+ \dots \theta_1^+|O(0)|\theta_1\ldots\theta_n\rangle \, ,\nonumber
\eea
where $\theta_i^+=\theta_i+i \epsilon_i$.  In this appendix, we will call the form factors like $\langle \theta_n^+ \dots \theta_1^+|O(0)|\theta_1\ldots\theta_m\rangle $ {\it regular} form factors, in order to distinguish them from the  {\it complete} form factors $\langle \theta_n \dots \theta_1|O(0)|\theta_1\ldots\theta_m\rangle $, that have also delta-like contributions. For $\epsilon_i$ finite these regular form factors are continuous functions but, despite their name, they can have a singular $\epsilon_i\to0$ limit, hence the need of the regularization procedure previously discussed. 
Let's firstly briefly anticipate the main steps of the proof:  after we expand the numerator of the l.h.s. of (\ref{A_starting}) (as done in (\ref{double})), we obtain
\bea
&&\!\!\!\!\!\!\!\!   \langle \tilde{B}|  {\cal O}(x,t)|\tilde{B}\rangle   = \sum_{n,l=0}^{+\infty} \frac 1 {n!\, l! }\int \frac{d\theta_1 \ldots d\theta_n}{(4\, \pi)^n} \, \frac{d\theta'_1 \ldots d\theta'_l}{(4\, \pi)^l} \exp\left[2\,i\,t(E_n(\theta)-E_l(\theta')) \right]\nonumber \cdot\\
&&\!\!\!\!\!\!\!\! \cdot\left[ \prod_{i=1}^n \overline{G}(\theta_i)\right]  \left[ \prod_{j=1}^l G(\theta'_j)\right]  \langle \theta_n,-\theta_n,\ldots \theta_1,-\theta_1|{\cal O}|-\theta'_1,\theta'_1,\ldots-\theta'_l,\theta'_l\rangle\,. \label{A_double}
\eea
Then, applying repeatedly the crossing relation (\ref{crossing}) to the matrix elements in ({\ref{A_double}), we arrive to an expression in which we can take the infinite time limit. In this limit, (\ref{A_double}) reduces to the r.h.s. of (\ref{A_starting}) times  the denominator of the l.h.s 
\be
{\cal Z}=\langle \tilde{B}|\tilde{B}\rangle =\langle \tilde{B}|e^{i{\cal H} t}e^{-i{\cal H} t} |\tilde{B}\rangle  =\sum_{n=0}^{+\infty} {\cal Z}_n\, , \label{A_Z}
\ee
where 
\bea
&&{\cal Z}_n= \frac 1 {n!^2} \int \frac{d\theta_1 \ldots d\theta_n}{(4\, \pi)^n} \, \frac{d\theta'_1 \ldots d\theta'_n}{(4\, \pi)^n}   \left[ \prod_{i=1}^n \overline{G}(\theta_i) G(\theta_i') \right] \cdot \\
&&\cdot\exp(2\,i\,t\,(E_n(\theta)-E_n(\theta')) \langle \theta_n,-\theta_n,\ldots \theta_1,-\theta_1|-\theta'_1,\theta'_1,\ldots-\theta'_n,\theta'_n\rangle\,, \nonumber 
\eea
thus completing the proof. 
\subsection*{Introductory remarks}
Here we would like to highlight some basic facts that we will find useful for our proof. First of all, we notice that, when we recast (\ref{A_double}) in terms of the regular form factors, the only terms that survive in the infinite time limit are the  time-independent ones. As a consequence, we can discard all the terms in the double sum in ({\ref{A_double}) where $n\ne l$. 
Moreover, from (\ref{crossing}) it is clear that the term $\langle0|{\cal O}|0\rangle$ in the r.h.s. of (\ref{A_starting}) is correct, so in the following we will not consider this contribution. 

In order to follow more easily our ideas, it may be useful to develop a diagrammatic representation. In fig. \ref{fig_basic} it is shown a {\em bra} state with $n$ pairs (a {\em ket} state can be introduced in a similar way). An inner product between a {\em bra} and a {\em ket} both made of $n$ pairs can be represented as the sum of all the possible ways to link together the particles of the ket with the particles of the bra, each link meaning a contraction between the corresponding particles. Of course, one should be careful about the permutation of particles and the corresponding $S$ matrix, but we will take care later of these details.
\begin{figure}[t]
\centering
\includegraphics[totalheight=0.2\textheight]{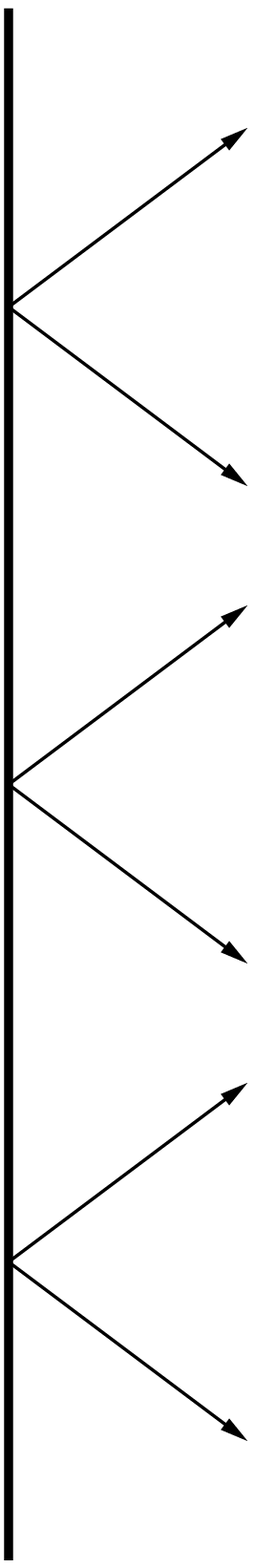}
\hfill
\includegraphics[totalheight=0.2\textheight]{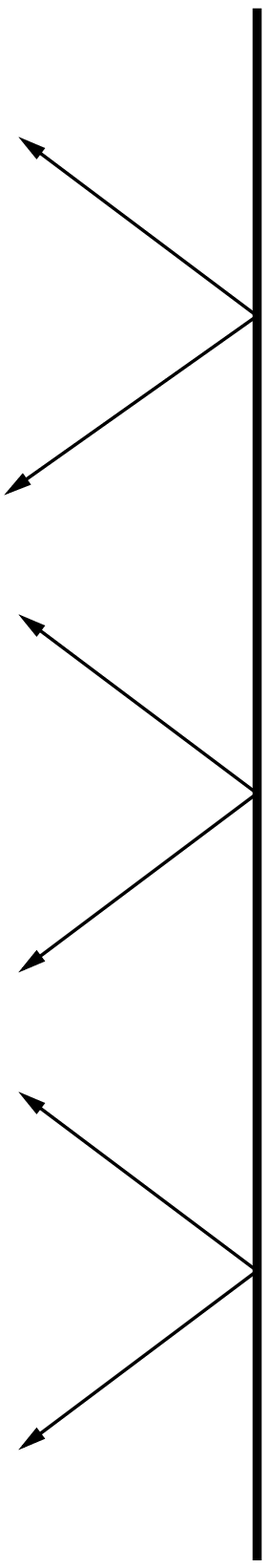}
\hfill
\includegraphics[totalheight=0.2\textheight]{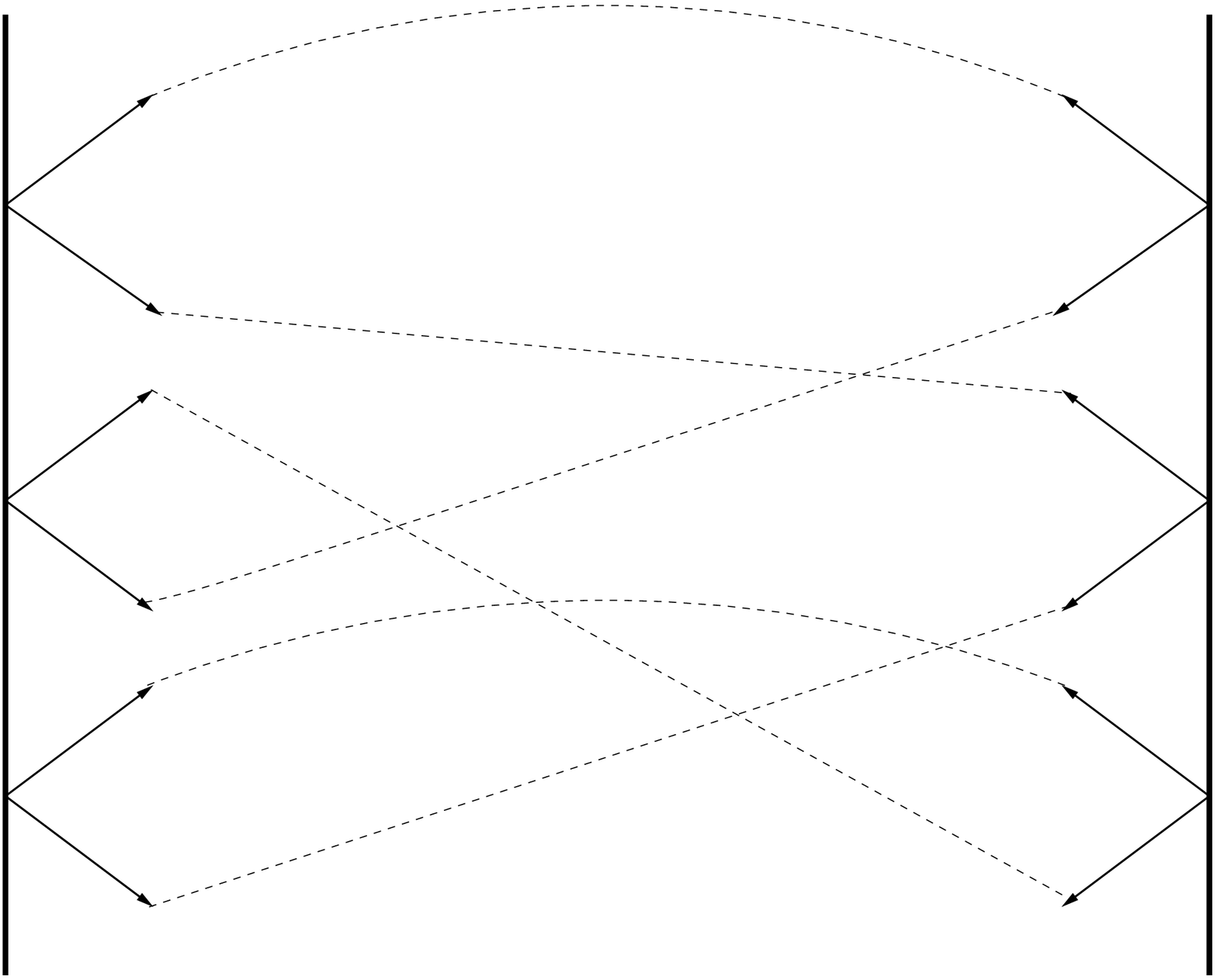}
\hfill
\includegraphics[totalheight=0.2\textheight]{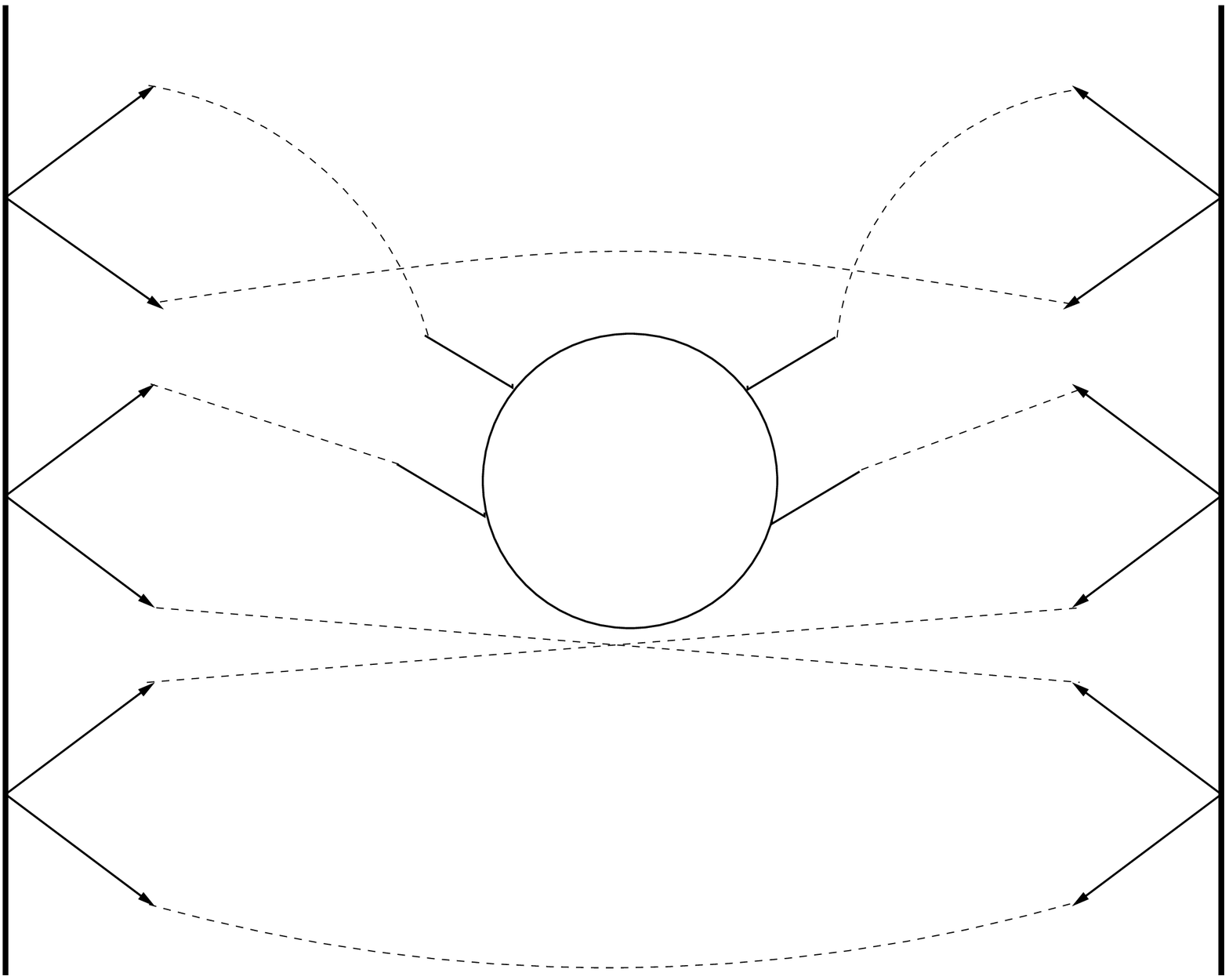} 
\caption{In this figure we present the building blocks of our diagrammatic representation. On the left, we can see a bra and a ket with $n=3$ Cooper pairs. In the middle, we see one of the diagrams that represent the inner product of the bra and the ket. The figure on the right shows a term proportional to the regular form factor $\langle\theta_2^+ ,\theta_1^+ |{\cal O}|\theta'_1,\theta'_2\rangle $. The circle in the middle stands for the operator $\cal O$, that has $2 r=4$ legs,  half connected to the bra and half to the ket.} \label{fig_basic}
\end{figure}
We are interested in the matrix elements of the operator $\cal O$ between states made of Cooper pairs. In particular, our aim is to reduce the full matrix element $\langle \theta_n,-\theta_n,\ldots \theta_1,-\theta_1|{\cal O}|-\theta'_1,\theta'_1,\ldots-\theta'_n,\theta'_n\rangle$ to the regular form factors $ \langle \theta_{i_r}^+,\ldots,\theta_{i_1}^+|{\cal O}|\theta'_{j_1},\ldots, \theta'_{j_r} \rangle$. These regular terms can be  diagrammatically represented as in fig. \ref{fig_basic}, where the operator ${\cal O}$ has $2r$ legs: $r$ of them are connected to the particles in the bra, while the other $r$ are connected to the particles in the ket. So, our combinatorial problem reduces to a problem in which we have to connect the legs of the operator to the bra/ket and the remaining particles together, according to (\ref{crossing}). 
In order to get the right combinatorial coefficients, we have to remember what follows.
\begin{itemize}
\item As we stated before, only the matrix elements with the same numbers of particles in the ket and in the bra give a non vanishing contribution in the long times limit. Therefor, the number of legs connected to the bra is always equal to the number of legs connected to the ket and, from now on, we will only specify the number of particles connected to the bra.

\item In principle, when we consider the matrix element between two states with $n$ pairs, i. e. $\langle \theta_n,-\theta_n,\ldots \theta_1,-\theta_1|{\cal O}|-\theta'_1,\theta'_1,\ldots-\theta'_n,\theta'_n\rangle $, we could expect to end up with a sum of regular terms with r legs linked to the bra, with $r\le 2 n$. However, it turns out  that  $r\le n$: if we connect r particles to the operator, we are left with $2n-r$ delta functions, and to suppress all the time dependencies,  we have to eliminate $n$ integration variables, hence $r\le n$.

\item Finally, if we link a particle to the operator, its pair partner {\it cannot}
be connected to $O$, otherwise their time dependence survives.
\end{itemize}
\subsection*{The disconnected terms}
In this subsection we will show how some contributions to (\ref{A_starting})  (the {\it disconnected terms}) cancel out with the denominator $\cal Z$ (\ref{A_Z}).
In order to understand this point, we analyze the matrix elements $\langle \theta_2,-\theta_2,\theta_1,-\theta_1|{\cal O}|-\theta'_1,\theta'_1,-\theta'_2,\theta'_2\rangle\ $. As it is shown in fig.~\ref{fig_example}, we have essentially two types of diagram.  Let's focus our attention on the second one:  it is clear that the contribution in the dotted box factorize from the integral with the form factors. Hence, if we have a set of particles that is completely disconnected from the operator $\cal O$ (this means that no one of these particles or their partners is connected to $\cal O$ or to a particle whose partner is connected to $\cal O$), its contribution factorizes.
\begin{figure}[t]
\centering
\includegraphics[totalheight=0.2\textheight]{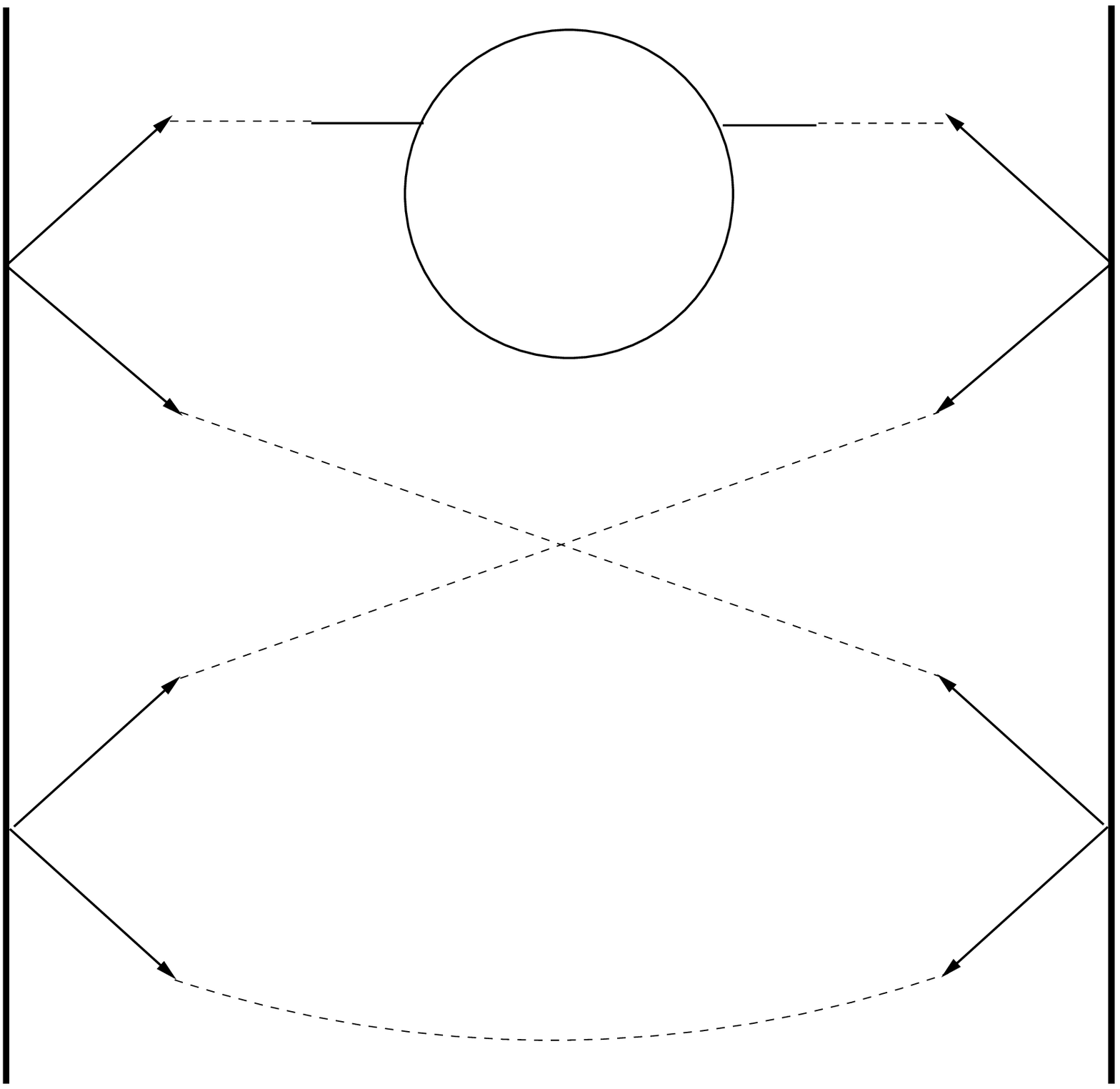}
\hfill
\includegraphics[totalheight=0.2\textheight]{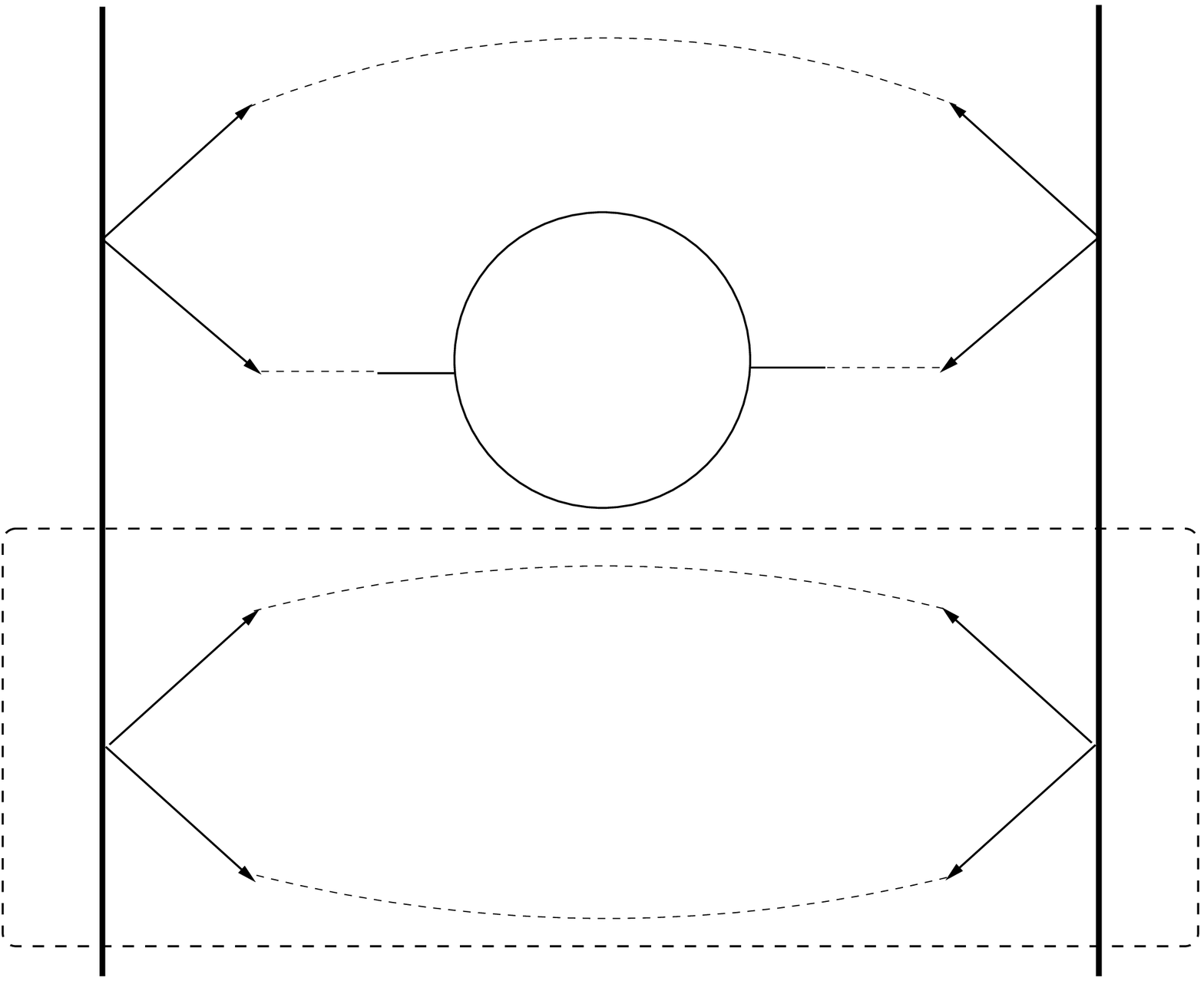}
\caption{Two different diagrams: in the left one all the particles are connected to the operator, while in the right one there is a contribution that factorizes.} \label{fig_example}
\end{figure}

What we want to show now is  that these disconnected pieces cancel out with the denominator $\cal Z$. 
Let us consider the term in (\ref{A_double}) with $n$ pairs in the bra (let's denote it as ${\cal O}_n$), and let's focus our attention on the contribution ${\cal O}_{[n,k]}$, such that $k$ pairs in the bra (as well as $k$ pairs in the ket) are disconnected and so only $n-k$ pairs in the bra are connected to $\cal O$. Of course, a similar term can also be obtained from ${\cal O}_{n-k}$, when no particle is disconnected. With our notation, this term can be written as ${\cal O}_{[n-k,0]}$ . If we are able to show that, for any $n$ and $k\le n$, 
\be
{\cal O}_{[n,k]}={\cal O}_{[n-k,0]}\, {\cal Z}_k\, , \label{A_disc}
\ee
then it follows that, when we sum over all n and $k\le n$, (\ref{A_double}) becomes
\be
\left( \sum_{m=0}^{+\infty} {\cal O}_{[m ,0]} \right) \, {\cal Z} \, ,
\ee
hence the disconnected pieces cancel out with ${\cal Z}$. 

The proof of (\ref{A_disc}) is actually quite simple. It is clear that the inner product of the $k$ disconnected pairs gives an integral proportional to ${\cal Z}_k$, as well as $ {\cal O}_{[n,k]}$ is clearly proportional to $ {\cal O}_{[n-k,0]}$. So, in order to complete the proof, we have only to check the proportionality constant. ${\cal O}_{[n,k]}$ has an overall coefficient $ \frac 1 {n!^2}$ from the expansion of the exponentials, while we can choose the k disconnected pairs in $\left( \frac{n!}{k!(n-k)!} \right)^2$ equivalent ways, since the creation operator of a pair commutes ($[Z^\dagger(-\theta)\, Z^\dagger(\theta), Z^\dagger(-\theta')\, Z^\dagger(\theta')]=0$). Nicely, $\frac 1 {(n-k)!^2}$ is the correct coefficient for ${\cal O}_{[n-k,0]}$ and $\frac 1 {k!^2}$ is the right one for ${\cal Z}_k $, thus concluding the proof of {\ref{A_disc}.
\subsection*{The last step}
Here we conclude our proof,  showing that $  {\cal O}_{[m ,0]}  $ has actually the right structure to give (\ref{A_starting}).
Let's call ${\cal O}_{\{m,r\}}$ the contribution where the operator has $r$ legs connected to the bra, the bra has $m$ pairs and no particle is disconnected from $\cal O$. We have already pointed out that for $t\to+\infty$
\be
{\cal O}_{[m ,0]}=\sum_{r=0}^{m} {\cal O}_{\{m,r\}} ,
\ee
since the sum is restricted to $r\le m$. So, in order to get (\ref{A_starting}), we need only to show that
\bea
&&{\cal O}_{\{m,r\}}=\frac 1 {r!}  \int\frac{d\theta_1 \dots d\theta_r} {(2 \pi)^r} \langle \theta_r^+ \dots \theta_1^+|O(0)|\theta_1\ldots\theta_r\rangle \cdot \label{A_step} \\
&&\cdot{\sum_{i_1,\ldots,i_r}}^\prime \left[S(0)^{i_1-1}( |G(\theta_1)|^2)^{i_1} \ldots S(0)^{i_r-1}( |G(\theta_r)|^2)^{i_r} \right] ,\nonumber
\eea
where the summation ${\sum_{i_1,\ldots,i_r}}^\prime$ is over all the positive integers $i_j$ such that $\sum_j i_j=m$.\\
In order to prove (\ref{A_step}), we need to be a little careful with the ordering of the particles and the labeling of the rapidities. However, if we exchange two particles, the contribution is same, since (as we already know) the pairs do commute while the exchange of two particles forming a pair is equivalent to change of the integrable variable $\theta\to-\theta$. This is a consequence of the symmetry of $G$ (\ref{cross_uni2}). 

Before concluding our proof of (\ref{A_starting}), we need to understand how to label the rapidity. We start with $2 m$ integration variables and the delta-functions reduce them to $r$. So, we use the convention to label the rapidities as in ({\ref{A_step}): we call $\theta_1$ the rapidity of the particle of the bra closest to ${\cal O}$ and we take advantage of delta functions in such a way that the rapidity of the particle in the ket nearest to $\cal O$  is also $\theta_1$, and so on.

We can now finally show that we obtain exactly the structure (\ref{A_step}). First of all, from our previous reasoning about the long time limit, it is clear that, performing all the possible contractions, for the term in the second row of 
(\ref{A_step}) we arrive to an expression as
\be
{\sum_{i_1,\ldots,i_r}}^\prime \left[c_1 (i_1) ( |G(\theta_1)|^2)^{i_1} \ldots c_r (i_r) ( |G(\theta_r)|^2)^{i_r} \right] ,
\ee
where $c_1 (i_1) \ldots c_r (i_r) $ are unknown constant. What we have to prove is that
\begin{enumerate}
\item The overall coefficient agrees with (\ref{A_step})
\item $c_j (i_j)=S(0)^{i_j-1}$.
\end{enumerate}
The first point  comes out from the combinatorial coefficient that takes in account all the equivalent way to link the particles and the operator. In ${\cal O}_{[m ,0]}$, we have an overall coefficient that is $\frac 1 {m!^2} \frac 1 {2^{2m}}$: the factorial comes from the exponentials while we get  the $\frac 1 2$ from the integration measure that is $\frac {d\theta}{4\pi}$ and  not $\frac  {d\theta}{2\pi}$. We can choose the $r$ particles in the bra (and the r in the ket) connected to ${\cal O}$ in 
\be
\left[\frac {2m \,2 (m-1)\ldots 2 (m-r-1)}{r!} \right]^2
\ee
ways.  We remind that if a particle is connected to $\cal O$, its pair companion cannot be directly connect to the operator, otherwise the contribution is time dependent hence it goes to zero for long times. We have to determine in how many ways we can connect the particle in the bra with rapidity $\theta_1$ to the others in order to have a term like $(|G(\theta_1)|^2)^{i_1} $. The answer is 
\be
\left[2(m-r)\,2(m-r-1)\ldots 2(m-r-i_1+1) \right]^2 r\,,
\ee
where the $r$ comes out from the $r$ equivalent ways to choose a particle in the ket connected to $\cal O$. If we repeat the same argument for all the particles, we end up with an overall coefficient that is
\be
\frac 1 {m!^2} \,\frac 1 {2^{2m}} \,\frac{2^{2r}} {r!^2} \,m!^2 \,2^{2(m-r)}\,  r!=\frac 1 {r!}\,,
\ee
in agreement with \ref{A_step}. 

Finally, we show that $c_j (i_j)=S(0)^{i_j-1}$. Since we know that every permutation of particles gives the same contribution, it is sufficient to show it only for one of the many equivalent ways to link particles. In particular,  we will consider the following way to separate the rapidities in two sets
\bea
&& \!\!\!\!\!\!\!\!\!\!\! \!\!\!\!\!\!\!\!\!\!\!  \!\!\!\!\!\!\!\!\!\!\!    \!\!\!\!\!      \langle\theta_m,-\theta_m,\ldots, \theta_1,-\theta_1 |{\cal O}| -\theta_1', \theta_1',\ldots -\theta_m', \theta_m,\rangle = S_{A A_1} S_{B B_1} \langle\theta_r^+,\ldots\theta_1^+ |{\cal O}| \theta_1',\ldots \theta_r'\rangle \cdot \nonumber\\ 
&& \!\!\!\!\!\!\!\!\!\!\!  \!\!\!\!\!\!\!\!\!\!\! \!\!\!\!\!\!\!\!\!\!\!  \!\!\!\!\!   \langle \theta_m,-\theta_m,\ldots,  \theta_{r+1},-\theta_{r+1},-\theta_r,\ldots-\theta_1|-\theta_1',\ldots -\theta_r', -\theta_{r+1}',\,\theta_{r+1}',\ldots,  -\theta_{m},'\theta_{m}'\rangle \,.\label{A_choice}
\eea
An useful trick is to remember that the contractions are such to have, at the end, $\theta_i=\theta_i'$ for $i=1,\ldots r$. When we impose this condition, we see that the S matrices in (\ref{A_choice}) reduces to the identity. Now, we want to contract $-\theta_1$ with $-\theta'_{r+1}$, obtaining a $\delta(\theta_1-\theta'_{r+1})$ and the desidered $|G(\theta_1)|^2$.  Then, we link $-\theta_{r+1}$ to $\theta'_{r+1}$. This contraction gives us a $\delta(\theta_1+\theta_{r+1})$. Finally, we commute $\theta_{r+1}$ (that now is equal to $-\theta_1$) with $-\theta_r \ldots -\theta_2$. In this way, we end up in a situation similar to the initial one. We have two pair less, a $|G(\theta_1)|^2$ overall and a huge product of $S$ matrices that comes from all the exchanges done. However, if we remind that at the end of our calculation we have  $\theta_i=\theta_i'$ for $i=1,\ldots r$, it is easy to see that this product of S matrices reduces to $S(0)$ . We can repeat this procedure until the overall coefficient is  $\left( S(0) |G(\theta_1)|^2 \right)^{i_1-1}$. Then we contract $-\theta_1$ with $ -\theta_1'$, obtaining another $|G(\theta_1)|^2 $ and we go on doing the same manipulations on $-\theta_2$. It is clear that at the end we obtain exactly \ref{A_step}.

\newpage
\bibliography{integrable}
\end{document}